\documentclass[review]{elsarticle}
\usepackage{subfig}
\usepackage{url} 
\usepackage{lineno, hyperref}
\usepackage{amsmath}
\usepackage{amssymb}
\usepackage{amsthm}
\usepackage[expert]{mathdesign}
\usepackage[mathscr]{eucal}
\hypersetup{ 
    colorlinks=true,       
    linkcolor=red,          
    citecolor=blue,        
    filecolor=magenta,      
    urlcolor=cyan           
}

\modulolinenumbers[1]

\journal{Journal of \LaTeX\ Templates}

\begin{document}
\nolinenumbers

\begin{frontmatter}

\title{Stochastic Integration of the Cahn-Hilliard Phase Field Equations}

\author[firstAdd]{Q. Yu}

\author[thirdAdd]{N. Julian}
\author[fourthAdd,fifthAdd]{J. Marian}

\author[firstAdd,secondAdd]{E. Martinez\corref{mycorrespondingauthor}}
\cortext[mycorrespondingauthor]{Corresponding author}

\address[firstAdd]{Department of Mechanical Engineering, Clemson University, Clemson, SC, United States}
\address[secondAdd]{Department of Materials Science and Engineering, Clemson University, Clemson University, Clemson, SC, United States}
\address[thirdAdd]{Los Alamos National Laboratory, Los Alamos, New Mexico, United States}
\address[fourthAdd]{Department of Mechanical and Aerospace Engineering, University of California, Los Angeles, California, United States}
\address[fifthAdd]{Department of Materials Science and Engineering, University of California, Los Angeles, California, United States}

\begin{abstract}
In this work we develop a stochastic algorithm to integrate the Cahn-Hilliard equations. The algorithm is based on Gillespie's stochastic simulation algorithm, also known as kinetic Monte Carlo. The deterministic integration of the phase field equations leads to the closest minimum of free energy and does not overcome free energy barriers. However, in the nucleation and growth regime of the phase diagram, free energy barriers need to be thermally overcome for the system to phase separate to reach equilibrium. We show in this work that the proposed stochastic integration algorithm indeed allows the system to overcome free energy barriers. We discuss the results in terms of fluctuation distributions, grid sizes and efficiency.  
\end{abstract}


\end{frontmatter}


\section{Introduction}\label{intro}

Through the last three decades,  the phase-field (PF) method has well demonstrated its unique outstanding benifit  over finite element method (FEM) in understanding materials' macroscopic properties through tracing microstructural evolutions during processing, especially under some uncertain circumstances such as arbitrarily complex interfaces between phases (free-boundary problems) whose evolution should be well taken care of. Basically, this method is characterized by treating interfaces as continuous fields, and simply obtaining the solution of interface evolution as well as other microstructural information through integrating a unique set of partial differential equations (i.e. Cahn-Hilliard\cite{Cahn1958} or Allen-Cahn\cite{refId0}) in the whole domain \cite{RN46}. Although the method has been applied to many different fields \cite{RN80, YANG20191, RN46, MCCUE201610}, the PF model as a tool is still facing the following challenges. First, nucleation process, as significant to microstructure selection, is hard to be addressed quantitatively and predictively with PF. Such problem arises from the fact that the formation of the nucleus of the second phase is always related to instabilities or disorders in local structures, which numerically is achieved by adding random noise/fluctuations. A few studies\cite{RN63, WANG19982983, Karma_1999, RN54} showed that the standard way of including noise is by adding a Langevin term, deduced from fluctuation-dissipation theorem, to the field equations. Typically this term is usually assumed to be $\delta-$correlated in space and time.  However, a direct comparison with atomistic results or classical nucleation theory is missing. Another challenge with PF is determining the correct size-discretization calibrated to physical parameters when upscaling PF from microscale to mesoscale and upwards.

Regarding the above two points, in this work we propose a new coarse-grained Spatially-Resolved Stochastic Phase Field model (SRSPF). This model is an extension of the SRSCD (Spatially-Resolved Stochastic Cluster Dynamics) method -- a mesoscale variant mean-field rate theory method -- that was used to study microstructure evolution during irradiation\cite{RN142797, RN142998}. The main advantage of our SRSPF method is that the intrinsic residence-time algorithm for sampling reactions naturally generates stochastic noise. We hypothesize that such noise will allow the system to overcome free energy barriers and will allow for the study of the nucleation and growth dynamics of the system. To show the capabilities of the SRSPF, we apply it to solve the precipitation process of a simple binary system, with atomic fraction as the only order parameter, and particle migration totally driven by chemical potential gradients. The free energy density of this example follows a double-well profile, and there is extra interface free energy depending on the second derivative of the solute concentration. Our main results show that first, the algorithm indeed overcomes free energy barriers.  Second, the nucleation rate obtained with the current model complies with classical nucleation theory (CNT). 


Finally, the discretization scheme in this mesoscale simulations leads to non-trivial effects on precipitation. Two theoretical rescaling rules of the parameters are derived and are then proved validated strategies for solving the dependence on the grid size.

 The paper is organized as follows. The phase field formulation on simple binary systems with a single conserved order parameter, and the simulation method, are introduced in Section~\ref{method}. Then, we present our results in Section~\ref{results}. We end with a discussion in Section~\ref{discussion} and the main conclusions in Section~\ref{conclusion}. 

\section{Theory and Method}\label{method}
\subsection{Phase field formulations}

\begin{figure}[htp]
\centering
  \includegraphics[width=0.8\linewidth]{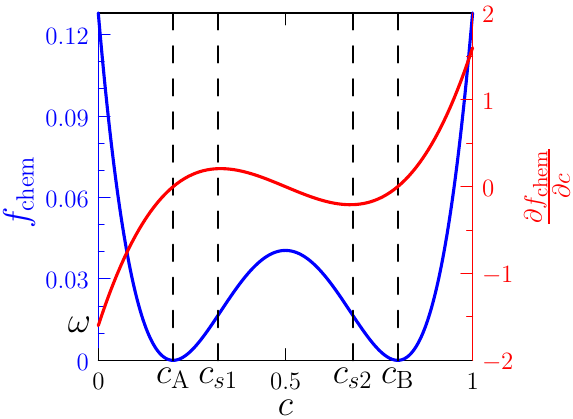}
  \caption{The free energy density $f_{\rm chem}$ (blue) and chemical potential $\frac{\partial f_{\rm chem}}{\partial c}$ (red) used in this work. $c_{\rm A}$ and $c_{\rm B}$ are equilibrium atomic fractions. $c_{s1}$ and $c_{s2}$ are boundary atomic fractions of spinodal decomposition.}
  \label{fig:freeenergy}
\end{figure}

The essential element for a typical phase field (PF) problem involves the construction of the functional of the total free energy, $F$, as a function of a set of non-conserved and conserved order parameters, respectively, $\phi_i$ and $c_i$, as in the following general expression \cite{RN47},
\begin{equation}
F = \int{f\left(\phi_1, \phi_2, ...,\phi_n, c_1, c_2...,c_n,\nabla\phi_1, \nabla\phi_2,...,\nabla\phi_n, \nabla c_1, \nabla c_2, ..., \nabla c_n, p, T, ...\right)dV}
\label{gen_fun}
\end{equation}
where $f$ represents the free energy density, which includes (1) a potential component that is governed by order parameters (e.g. $\phi_i$ and $c_i$), (2) an energic cost term, correlated with gradient terms (e.g. $\nabla \phi_i$ and $\nabla c_i$), due to interfaces, (3) other local state variables such as temperature $T$, pressure $p$, stress/strain, and/or external stimuli. For each PF problem expressed by eq.\eqref{gen_fun}, conserved and/or non-conserved variables are used to represent the microstructure of the system, with $c_i$ defined as physically measurable quantities, such as constituent concentrations or the relative density, while $\phi_i$ refers to structural features such as local crystallinity \cite{RN46, RN62}. Both $\phi_i$ and $c_i$ evolve with time $t$ to lower the total free energy $F$, such that a set of governing equations are obtained,
\begin{align}
\frac{\partial c_i}{\partial t} = \nabla \cdot \left(M_{c_i}\nabla\frac{\delta F}{\delta c_i}\right) = \nabla \cdot \left(M_{c_i}\nabla \left[ \frac{\partial f}{\partial c_i} - \nabla \cdot \frac{\partial f}{\partial \nabla c_i}\right]\right) \label{C-H}\\
\frac{\partial \phi_i}{\partial t} = -M_{\phi_i}\nabla\frac{\delta F}{\delta \phi_i} =  -M_{\phi_i} \left[ \frac{\partial f}{\partial \phi_i} - \nabla \cdot \frac{\partial f}{\partial \nabla \phi_i}\right] \label{A-C}
\end{align}
well-known as, respectively, Cahn-Hilliard \cite{Cahn1958} and Allen-Cahn \cite{refId0} equations with mobility coefficients $M_{c_i}$ and $M_{\phi_i}$. Primarily, the main difference of a given PF model to another lies in the group of order parameters to be chosen,  and the various contributions of each paramenter to $F$. In this paper, we aim at proposing a stochastic algorithm to integrate the phase field equations, which produces statistical noise. To validate the algorithm we consider a simple binary system with solute atomic fraction, $c$, as the only order parameter. Hence our total free energy functional $F$ is expressed as \cite{RN64},
\begin{equation}
F= \int_V\left(f_{\rm chem}\left(c\right)+ \frac{\kappa}{2}|\nabla c|^2\right)dV
\label{this_fun}
\end{equation}
where $f_{\rm chem}$ is the chemical free energy density and $\kappa$ is the gradient energy coefficient. Here, we define $f_{\rm chem}$ with a simple polynomial form,
\begin{equation}
f_{\rm chem}\left(c\right) = \rho_s \left(c-c_{\rm A}\right)^2\left(c_{\rm B}-c\right)^2
\label{f_density}
\end{equation}
such that $f_{\rm chem}$ shows a symmetric double-well profile with local minima at $c_{\rm A}$ and $c_{\rm B}$, corresponding to the equilibrium atomic fractions of phase $\rm A$ and phase $\rm B$, and $\rho_s$ controls the height of the double-well barrier. Then Cahn-Hilliard equation (eq. \eqref{this_fun}) is to be expressed as,
\begin{equation}
\frac{\partial c}{\partial t} =\nabla \cdot \left ( M\nabla \mu \right)= \nabla \cdot \left \{ M\nabla\left(\frac{\partial f_{\rm chem}}{\partial c} - \kappa \nabla^2c\right)\right\}
\label{C_H}
\end{equation}
where $\mu = \frac{\partial f_{\rm chem}}{\partial c} - \kappa \nabla^2c$ represents the solute chemical potential, and $M$ is the solute mobility. Example profiles for $f_{\rm chem}$ and $\frac{\partial f_{\rm chem}}{\partial c}$ used in this work are shown in Fig.~\ref{fig:freeenergy}, with values of parameters listed in Table~\ref{tab:msp}. Note that here we chose our formulation and parameters close to the benchmark problem provided in Ref.\cite{RN64} in which none of these values is actually related to real material's properties and should be regarded as dimensionless. In Table~\ref{tab:msp} two sets of parameters are listed, relating to two variant free energy density functions are used in this work. In section~\ref{results} most of the results are obtained for parameters \{$c_{\rm A}$,  $c_{\rm B}$, $c_{s1}$, $c_{s2}$\} with only those in section~\ref{e_kappa} done by set \{$c_{\rm A}'$,  $c_{\rm B}'$, $c_{s1}'$, $c_{s2}'$\}.

\begin{table}[htp!]
\caption{\label{tab:msp} Phase field parameters employed in this work.}
\centering
\footnotesize
\begin{tabular}{p{4.5cm}c c}
\hline
\hline
Parameter Name: &Symbol&Value\\
\hline
equilibrium atomic fraction & $c_{\rm A}$,  $c_{\rm B}$ & $0.2$, $0.8$ \\
&$c_{\rm A}'$,  $c_{\rm B}'$&$0.01$, $0.99$ \\
spinodal boundary & $c_{s1}$, $c_{s2}$ & $0.32$, $0.68$\\
&$c_{s1}'$, $c_{s2}'$&$0.22$, $0.78$\\
mobility & $M$ & $5.0$\\
reference free energy density & $\omega$ & -8.0\\
gradient energy coefficient & $\kappa$ & $3.0$\\
shape constant & $\rho_s$ & $5$ \\
atomic volume & $\Omega_a$ & $1$\\
\hline
\hline
\end{tabular}
\end{table}

\subsection{The kinetic Monte Carlo algorithm}

\begin{figure}
\centering
\includegraphics[width = 0.50\linewidth]{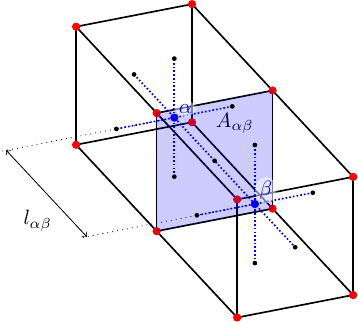}
\caption{Schematic diagram of two volume elements of a 3D space discretization used to calculate spatial gradients within SRSCD.\label{stencil}}
\end{figure}

One of the natural ways to account for fluctuations in atomic transport is by using the kinetic Monte Carlo (KMC) algorithm, in which a group of state-to-state transitions are sampled from their transition rates ($r_i $) and the time is sampled from an exponential distribution. Gillespie\cite{RN142622} derived the conditional probability distribution for the system to evolve from state to state and devised an algorithm to sample such distribution. Here we will use the direct method to sample escaping times and next events, and hence, obtaining one realization of the system dynamics. The escaping time follows an exponential distribution and can be drawn using the following expression,
\begin{equation}
dt = -\frac{1}{R_{\rm{tot}}}\ln \xi
\label{residence}
\end{equation}
where $R_{\rm{tot}} = \sum_i^n r_i$ is the sum of the rates of every possible event in the system, and $\xi \in [0,1)$ is a random number drawn from a uniform distribution. In the present work, we follow the Spatially Resolved Stochastic Cluster Dynamics (SRSCD) model \cite{RN142998, RN142797} to integrate the governing equations. The SRSCD is a stochastic variant of the mean field rate theory method, alternative to the standard ODE-based implementations, that eliminates the need to solve exceedingly large sets of ODEs and relies instead on sparse stochastic sampling from the underlying kinetic master equation \cite{RN22,RN20}. The spatial resolution of SRSCD is established by homogeneously subdividing the simulation bulk into $K$ pieces of domains/elements, such that the time evolution of $C_i^{\alpha}$, volume concentration of species $i$ at the current element $\alpha$, is governed by,
\begin{equation}\label{mfrt}
\frac{dC^{\alpha}_i}{dt}=\nabla \left(D_i \cdot \nabla C^{\alpha}_i\right)+ g_i+\sum_q\left( \sum_j s_{jq}C^{\alpha}_j-s_{iq}C^{\alpha}_i+\right)+\sum_{j}\left[\left(\sum_{h}k_{jh}C^{\alpha}_h-k_{ij}C^{\alpha}_i\right)C^{\alpha}_j\right]
\end{equation} 
where the first term on the right hand side of Eq.~\eqref{mfrt} is simply fickian's term of particle transport between neighboring volume elements. The set $\{{g}, {s}, {k}\}$ represents the reaction rates of $0^{th}$ (insertion), $1^{st}$ (thermal dissociation, annihilation at sinks), and $2^{nd}$ (binary reactions) order kinetic processes. In this work, the only reaction considered is solute migration, so those terms involving $\{{g}, {s}, {k}\}$ in Eq.~\eqref{mfrt} are discarded. Fick's equation is an approximation to Onsager's transport equations for non-interacting particles. In Onsager's formulation, particle transportation is driven by chemical potential gradients, so the first term of Eq.~\eqref{mfrt} is then replaced as, 
\begin{equation}\label{masterPFd}
\frac{dC^{\alpha}_i}{dt} = \nabla \left(M \cdot \nabla \mu\right)
\end{equation}
further, applying the divergence theorem to Eq.\eqref{masterPFd},
\begin{equation}\label{masterPFdd}
\int_\Omega \frac{dC_i}{dt}d\Omega = -\oint_{S} \left(M \cdot \nabla \mu\right)\vec n dS 
\end{equation}
and it eventually comes to the master equation in discrete form,
\begin{equation}\label{masterPFe}
\frac{dN_i^{\alpha}}{dt} = M\sum_{\beta}A_{\alpha\beta}\frac{\mu_i^{\beta}-\mu_i^{\alpha}}{l_{\alpha\beta}}
\end{equation}
where superscript $\beta$ referred to a neighboring element of $\alpha$ (see Fig.~\ref{stencil}).

For the flux of mass to be able to go uphill, i.e., in the direction opposite to the gradient of the chemical potential and hence consider important concentration fluctuations, we renormalize Eq.~\eqref{masterPFe} introducing a reference chemical potential $\omega$,


\begin{equation}\label{masterPF}
\frac{dN_i^{\alpha}}{dt} = -\sum_\beta \frac{M A_{\alpha\beta}}{l_{\alpha\beta}}\left(\mu^{\alpha}-\omega\right)+\sum_\beta \frac{M A_{\alpha\beta}}{l_{\alpha\beta}}\left(\mu^{\beta}-\omega\right)
\end{equation}
where $\omega \leq \mu_{\rm{min}}$. This renormalization will imply forward and backward mass fluxes among elements in the system even for flat concentration profiles. Each term on the right-hand-side represents a rate that can be sampled following the KMC algorithm to evolve the system in time as described in Ref.~\cite{RN22}.
 
\section{Results}\label{results}

Our results describe precipitation processes starting from a flat elemental concentration ($c_0$). $c_0$ is set outside the spinodal decomposition domain but inside the miscibility gap (i.e. $c_{\rm A} < c_0 < c_{s1}$), for deliberately making the SRSPF model to overcome free energy barriers solely by its stochastic noise. Every case in the figures below is the average result of five independent simulations with error bars denoting the standard error. In the following, we display the results from our model by showing the effects of some parameters, such as elemental size ($A_{\alpha\beta}$, $l_{\alpha\beta}$ and $\Omega$), $\omega$ and $\kappa$, on the precipitation dynamics. 

\subsection{Effect of $\omega$}
The physical meaning of $\omega$ is already mentioned above with a value equal to $\frac{\partial f_{\rm chem}}{\partial c}\left(0\right)$ or below. With such renormalization value, mass flux becomes possible under negative chemical potential gradients, which is a source of stochastic noise. To understand how $\omega$ affects the kinetics of precipitation, we analyze the process the system takes towards equilibrium. We study the evolution of the volume fraction of phase $\rm{B}$,  $\Phi_{\rm{B}}$, computed in SRSPF by the ratio of current number of elements with composition of phase B ($K_{\rm{B}}$) to the total number of elements ($K$), which at equilibrium should satisfy the lever rule,
\begin{equation}\label{equilibrium}
\Phi_{\rm{B}}^{eq} = \frac{K_{\rm{B}}}{K} = \frac{c_0-c_{\rm{A}}}{c_{\rm{B}}-c_{\rm{A}}}
\end{equation}   
Strictly speaking, an element can be regarded as `phase $\rm{B}$' only if its current concentration is exactly $c_{\rm{B}}$. However, owing to the stochastic nature of our model, we consider phase $\rm{B}$ as those elements having concentration $c \ge \frac{c_A+c_B}{2}$. 

\begin{figure}[ht!]
\centering
\subfloat[]{\includegraphics[width=.48\textwidth]{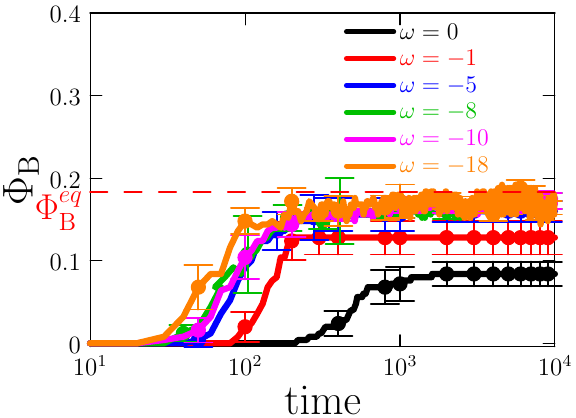}\label{volumeF_omega}}\quad
\subfloat[]{\includegraphics[width=.47\textwidth]{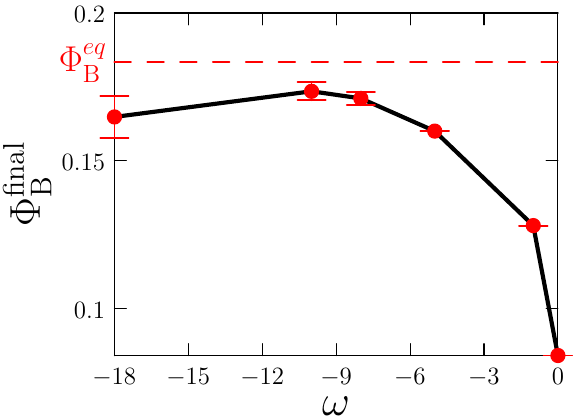}\label{finalVF}}\quad
\subfloat[]{\includegraphics[width=.48\textwidth]{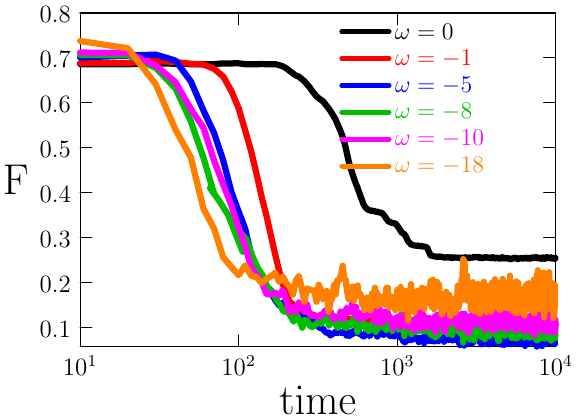}\label{FE_omega}}\quad
\subfloat[]{\includegraphics[width=.47\textwidth]{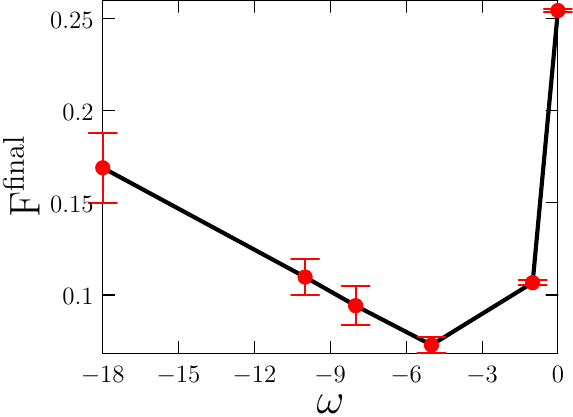}\label{finalFE}}\\
\caption{Precipitation process for different values of $\omega$. The initial concentration is $c_0 = 0.31$. (a) and (c) are time evolution of the volume fraction of phase $\rm{B}$ ($\Phi_{\rm{B}}$), and total the free energy ($\rm{F}$), respectively. (b) and (d) are the average values of both of the quantities from the last 10$^3$ time units.}
\label{omega}
\end{figure}

Figures~\ref{volumeF_omega} and~\ref{FE_omega} show the evolution of $\Phi_{\rm{B}}$ and $F$ during precipitation for different values of $\omega$ up to 10$^4$ time units, from which three stages can be distinguished: (1) pre-nucleation stage with very low $\Phi_{\rm{B}}$ ($\approx 0$) but high $F$; (2) nucleating stage having exponentially increasing $\Phi_{\rm{B}}$ and decreasing $F$; (3) post-nucleation stage where both $\Phi_{\rm{B}}$ and $F$ converge to a stable/meta-stable state. To understand the contribution of $\omega$ to the equilibrium state, we further display the average $\Phi_{\rm{B}}$ and $F$ from the last one thousand time units in Figs.~\ref{finalVF} and~\ref{finalFE}. For $\omega$ in the range of approximately -15 to -5, the volume fraction of phase $\rm{B}$ reaches a value close to equilibrium. The decrease in $\Phi_{\rm{B}}$ and increase in $F$ for low $\omega$ (large in absolute value) is possibly due to the anharmonicity of the free energy function. As a low value of $\omega$ induces larger noise, anharmonicity leads to a change in average. We will be using $\omega =-8.0$ throughout the rest of this work.

\subsection{Effect of $\kappa$}\label{e_kappa}

\begin{figure}[ht!]
\centering
\subfloat[]{\includegraphics[width=.48\textwidth]{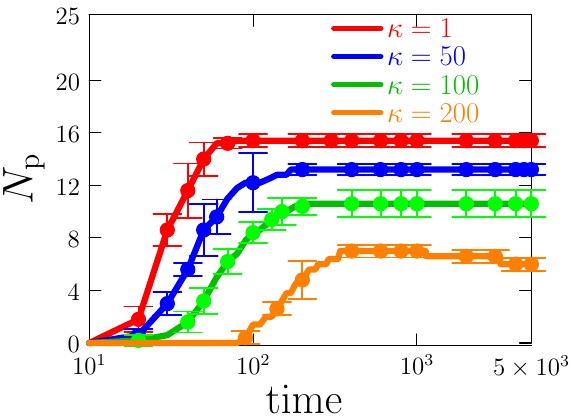}\label{volumeF_kappa}}\quad
\subfloat[]{\includegraphics[width=.49\textwidth]{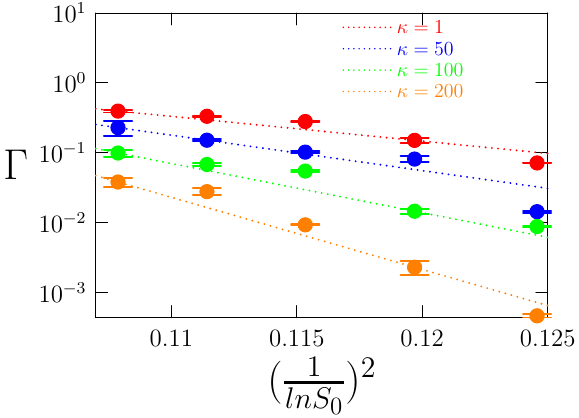}\label{rate_super}}\\
\caption{(a)Time evolution of $N_{\rm P}$ at $S_0 = 0.21$  (b) $\ln \left(\Gamma \right)$ v.s. $\left(\frac{1}{\ln S_0} \right)^2$ profiles at various $\kappa$. In (b), the solid circle points are measured values of simulations at $S_0 =21, 20, 19, 18$ and 17; the dashed lines are the fitting curves (see Eq.~\eqref{nucl_rate}) of which the slope is actually `$-\theta \left(\kappa \right)$'.}.
\label{kappa}
\end{figure}

\begin{figure}[ht!]
\centering
\includegraphics[width=.7\textwidth]{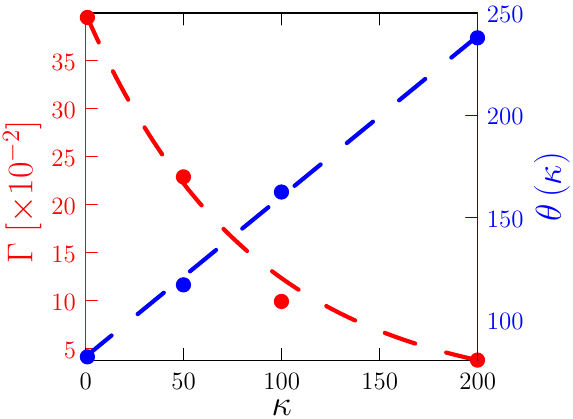}
\caption{Values of $\Gamma$(red) in Fig.\ref{volumeF_kappa} and $\theta \left(\kappa \right)$(blue) in Fig.~\ref{rate_super}. The dashed lines are the fitting curves.}
\label{nr_s0}
\end{figure}

The interface free energy term which is pre-factored by $\kappa$ slightly modifies the local free energy from equilibrium (Gibbs-Thompson effect), such that with the increase of $\kappa$, the energetic cost of generating interface increases the free energy barrier for nucleation leading to a slower nucleation rate. Such effect of interface free energy is well understood from classical nucleation theories \cite{PhysRevB.62.203}, which show that the steady-state nucleation rate $\Gamma$ is generally expressed by,
\begin{equation}\label{nucl_rate}
\Gamma = \Gamma_0 \exp \left( -\frac{\theta \left(\kappa\right)}{\left(\ln S_0\right)^2}\right)
\end{equation}
 where $\theta \left(\kappa \right)$ is a power function of the interface free energy, and thus in our case a power function of $\kappa$. $S_0$ represents supersaturation and equals simply $S_0 = \frac{c_0}{c_{\rm{A}}}$. The time evolution of the number of nuclei $N_{\rm P}$ under various $\kappa$ at $S_0 =21$ is presented in Fig.~\ref{volumeF_kappa}. To quantify the nucleation rate, the slope of the steepest segment of each curve is measured in Fig.~\ref{volumeF_kappa}, and is summarized in Fig.~\ref{nr_s0} (the red dashed line). As expected, there is a drop in $\Gamma$ from $0.39$ at $\kappa = 1$ to $0.038$ at $\kappa = 200$. 

Finally, results for $S_0$ equal to 20, 19, 18 and 17 were also obtained, leading to the relation of $\Gamma$ versus $\left(\frac{1}{\ln S_0} \right)^2$ shown in Fig.~\ref{rate_super}. In this figure, a general exponentially decreasing trend (note logarithmic scale in the figure) of $\Gamma$ with decreasing $S_0$ is observed (solid points) for each $\kappa$ case, which implies that the initial concentration not only determines $\Phi_{\rm{B}}^{eq}$, but also has an impact on the precipitation rate, as predicted by classical nucleation theory. The fitted $\theta \left( \kappa \right)$ is displayed in Fig.~\ref{nr_s0} (blue) where it almost linearly increases from 81.6 at $\kappa = 1$ to 237.8 at $\kappa = 200$. The figure (Fig.~\ref{nr_s0}) also shows the rate depending on $\kappa$ following an exponential decay.

\subsection{Rescaling rules and size discretization effects}\label{rescale}

\begin{figure}[ht!]
\centering
\subfloat[]{\includegraphics[width=.31\textwidth]{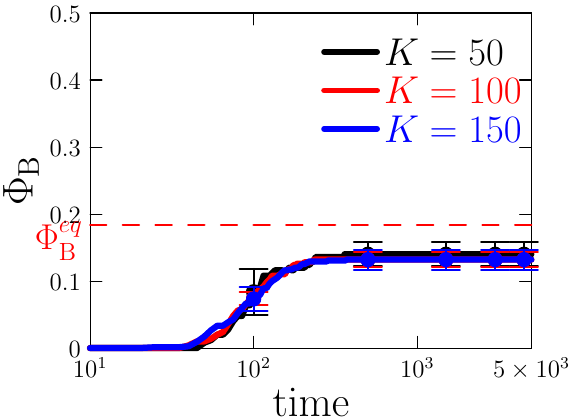}\label{volumeF4}}\quad
\subfloat[]{\includegraphics[width=.31\textwidth]{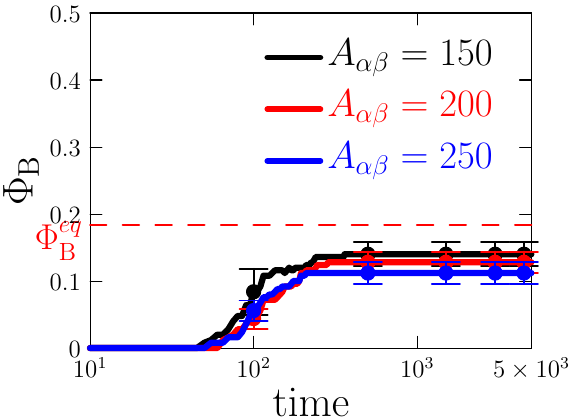}\label{volumeF8A}}\quad
\subfloat[]{\includegraphics[width=.31\textwidth]{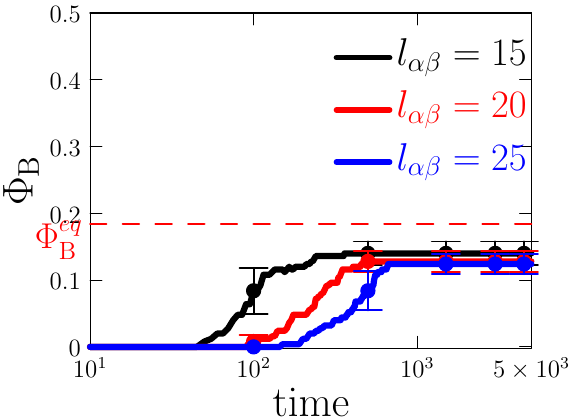}\label{volumeF5A}}
\caption{Stochastic precipitation process for an initial concentration $c_0=0.31$ obtained varying (a) $K$, (b) $A_{\alpha\beta}$, (c) $l_{\alpha\beta}$. In these figures, $\kappa = 0$, $K = 50$.}
\label{KAL}
\end{figure}

The kinetics of the precipitation process should not be influenced by any factors other than the physical parameters (Eq.\eqref{C_H}). However, as suggested in Eq.~\eqref{masterPF}, the discretization parameters such as $K$, $A_{\alpha\beta}$ and $l_{\alpha\beta}$ are involved in the rate equation and might affect the evolution of $\Phi_{B}$, $F$, and any other physical observable. In Fig.~\ref{KAL}, we sequentially show the effect of $K$, $A_{\alpha\beta}$ and $l_{\alpha\beta}$ on precipitation. It is interesting to see that $K$ does not affect the precipitation kinetics and the effect of $A_{\alpha\beta}$ seems small, while the effect of $l_{\alpha\beta}$ appeared to be significant. We analyze such effects of the discretization in the following.

Different size discretizations in the SRSPF model induce two main effects to the system. One was already shown in Eq.~\eqref{masterPF}: the event rates $r_i$ depend on the term $\frac{A_{\alpha\beta}}{l_{\alpha\beta}}$ which relates to the kinetics of phase transitions. The second effect of discretization is in the magnitude of the fluctuations of the order parameter per event, $\Delta C^{\alpha} =1/{\Omega} $. Thermally activated processes rely on fluctuations to overcome free energy barriers. Hence, different discretizations, and hence fluctuations, might lead to different nucleation kinetics. Therefore, understanding these effects is critical to be able to mitigate them. The first effect can be resolved by rescaling parameters such as $M$ and $\kappa$ (there is a second derivative of concentration in the expression for $\mu$); and the second effect can be resolved by fixing the elemental volume.

In this work, we propose two equivalent renormalization rules. First, extracting and expanding size-related terms in Eq.~\eqref{masterPF} gives
\begin{equation}\label{exp1}
\frac{MA_{\alpha\beta}}{l_{\alpha\beta}}\mu^{\alpha} = \frac{MA_{\alpha\beta}}{l_{\alpha\beta}}\left[2\rho_s \left(c^{\alpha}-c_{\rm A}\right)\left(c_{\rm B}-c^{\alpha}\right)\left(1-2c^{\alpha}\right)-\kappa\frac{c^{\alpha-1}-2c^{\alpha}+c^{\alpha+1}}{l_{\alpha\beta}^2} \right]
\end{equation}
The first renormalization rule (RR1) can be expressed as
\begin{equation}\label{RR1}
\begin{cases}
M' &= M\frac{l_{\alpha\beta}}{A_{\alpha\beta}}\\
\kappa' &= \kappa\frac{l_{\alpha\beta}^2}{l_0^2}\\
\Omega_{\alpha} = l_{\alpha\beta}A_{\alpha\beta} &\equiv \rm{const.}
\end{cases}
\end{equation}
where $l_0$ is an arbitrary parameter that can be optimized for computational efficiency. The second renormalization rule (RR2) is obtained by expressing Eq.~\eqref{exp1} in terms of the particle number ($N^{\alpha}$). That is,
\begin{equation}\label{exp2}
\begin{split}
\frac{MA_{\alpha\beta}}{l_{\alpha\beta}}\mu^{\alpha} &= MA_{\alpha\beta}\left[\frac{2\rho_s(N_{\rm A}-N^{\alpha})(N_{\rm B}-N^{\alpha})(\Omega_{\alpha}-2N^{\alpha})}{l_{\alpha\beta}\Omega_{\alpha}^3}-\kappa\frac{N^{\alpha-1}-2N^{\alpha}+N^{\alpha+1}}{\Omega_{\alpha} l_{\alpha\beta}^2}\right]\\
& = \frac{M}{l_{\alpha\beta}^2}\left[\frac{2\rho_s(N_{\rm A}-N^{\alpha})(N_{\rm B}-N^{\alpha})(\Omega_{\alpha}-2N^{\alpha})}{\Omega_{\alpha}^2}-\kappa\frac{N^{\alpha-1}-2N^{\alpha}+N^{\alpha+1}}{l_{\alpha\beta}}\right]
\end{split}
\end{equation}

which leads to the renormalized parameters of the RR2 approach as,
\begin{equation}\label{RR2}
\begin{cases}
M' &= M\frac{l_{\alpha\beta}^2}{l_0^2}\\
\kappa' &= \kappa\frac{l_{\alpha\beta}}{l_0} \\
\Omega_{\alpha} = l_{\alpha\beta}A_{\alpha\beta} &\equiv \rm{const.}
\end{cases}
\end{equation}
In Eq.~\eqref{exp2}, the fact that the first term depends on $\Omega_{\alpha}$ implies the need for keeping the elemental volume constant to maintain constant rates and equal fluctuations. The dependence on the elemental volume comes from the fact that the homogeneous term of the chemical potential depends on the local volume. 

\begin{figure}[ht!]
\centering
\subfloat[]{\includegraphics[width=.31\textwidth]{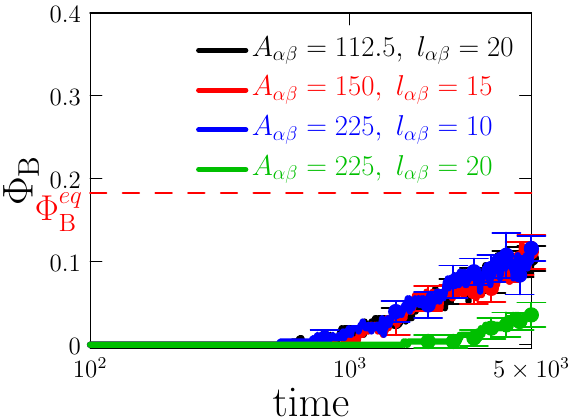}\label{volumeFA}}\quad
\subfloat[]{\includegraphics[width=.31\textwidth]{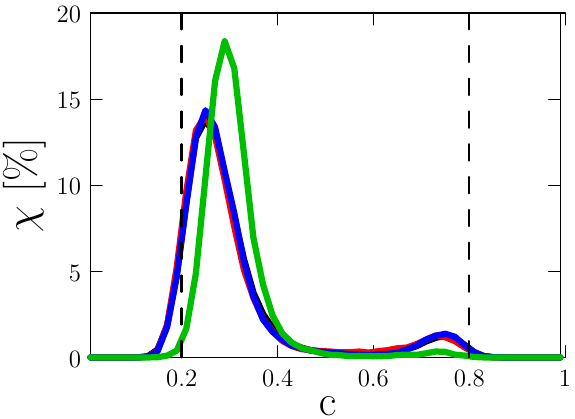}\label{frequencyFA}}\quad
\subfloat[]{\includegraphics[width=.31\textwidth]{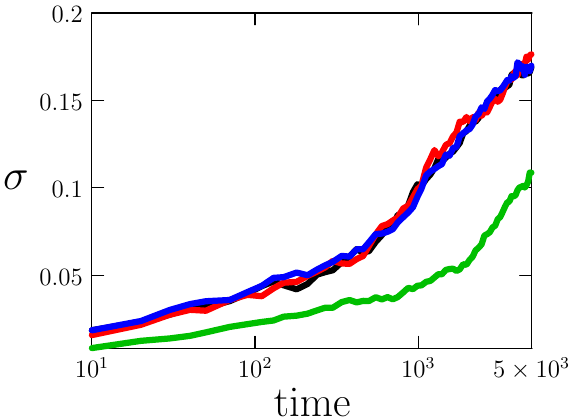}\label{timeEFA}}
\caption{ RR1 for an initial concentration $c_0=0.31$. In these figures, $\kappa = 50$, $K = 50$, $l_0 = 15$ and $\omega = -8.0$.}
\label{Rule1}
\end{figure}

\begin{figure}[ht!]
\centering
\subfloat[]{\includegraphics[width=.31\textwidth]{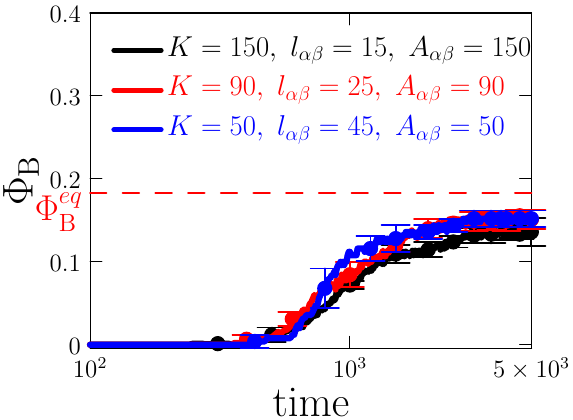}\label{volumeFR2}}\quad
\subfloat[]{\includegraphics[width=.31\textwidth]{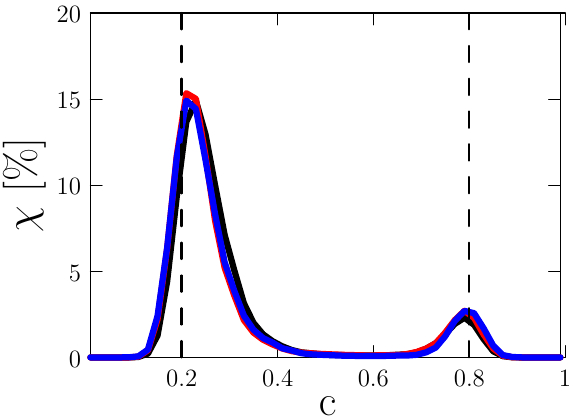}\label{frequencyR2}}\quad
\subfloat[]{\includegraphics[width=.31\textwidth]{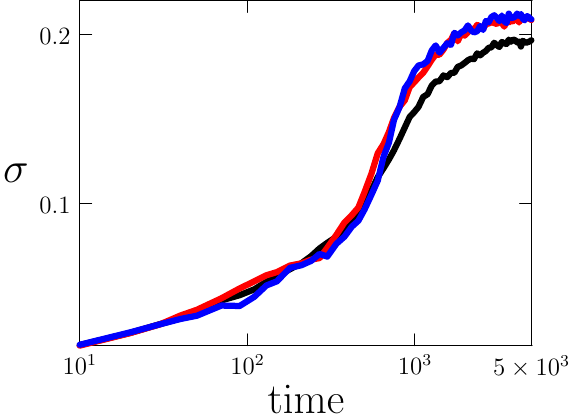}\label{timeER2}}
\caption{Stochastic precipitation under RR2 for an initial concentration $c_0=0.31$. In these figures, total length of the system ($Kl_{\alpha\beta}$) is fixed and is subdivided in three variant ways. Other parameters are $\kappa = 50$, $l_0 = 50$ and $\omega = -8.0$.}
\label{Rule2}
\end{figure}

Figures ~\ref{Rule1} and~\ref{Rule2} show the precipitation processes with renormalization RR1 and RR2, respectively. We compare in Fig ~\ref{Rule1}(a) four different cases, three of them conserve volume while one (green line $A_{\alpha\beta} = 225,\ l_{\alpha\beta}= 20$) does not. As a result, the three cases conserving volume follow a similar kinetic evolution, with the volume fraction of all cases within error bars. However, the case with a different volume follows a different evolution highlighting the need of the renormalization scheme for the evolution not to depend on discretization parameters. 

To further compare the similarity among cases, we also show the concentration distribution ($\chi$ [\%]) in Figs.~\ref{frequencyFA} and ~\ref{frequencyR2}, and the standard deviation of the solute concentration with respect to the nominal concentration $c_0$ ($\sigma\left(t\right)$) in Figs.~\ref{timeEFA} and~\ref{timeER2}. Here $\chi$ is obtained by counting the number of volume elements within a composition range, with a concentration interval of $\Delta c=0.02$,
\begin{equation}\label{chi}
\chi^{b_i} = \frac{\Lambda^{b_i}}{\Lambda^{\rm{tot}}}\times 100
\end{equation}  
where $b_i$ represents a concentration interval $[c_i, c_{i+1})$, $\Lambda^{b_i}$ and $\Lambda^{\rm{tot}}$ are the number of the elemental counts with solute composition lying in $b_i$, and total number of elemental counts, respectively. $\chi$ identifies how close the system is to its equilibrium state. We observe that after $5,000$ time units the concentration distribution is similar for the three cases following RR1 but is different for the cases that does not.

Moreover, to understand the role of fluctuations we define $\sigma\left(t\right)$ as,
\begin{equation}\label{std}
\sigma\left(t\right)  = \sqrt{\frac{\sum_{\alpha}\left(c^{\alpha}(t) - c_0\right)^2}{K}}.
\end{equation}
$\sigma\left(t\right)$ is the standard deviation of the concentration distribution and shows how far the system has been driven away from its initial homogeneous state by the driving force $\nabla \mu$. We observe in Fig.~\ref{Rule1} that the case with a different volume leads to different $\chi$ and $\sigma$ profiles comparing to the other three cases, which suggests the need of to conserve elemental volume. $\sigma$ seems to converge as the volume fraction reaches steady state.



\begin{figure}[ht!]
\centering
\subfloat[]{\includegraphics[width=.48\textwidth]{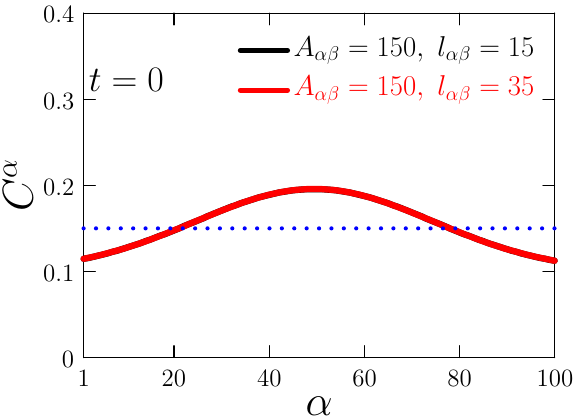}\label{hd0}}\quad
\subfloat[]{\includegraphics[width=.48\textwidth]{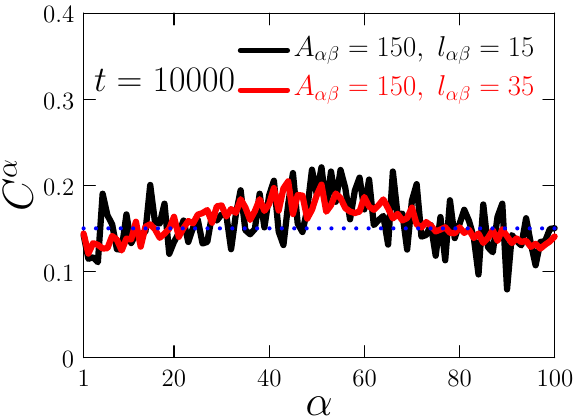}\label{hd1000}}\quad
\subfloat[]{\includegraphics[width=.48\textwidth]{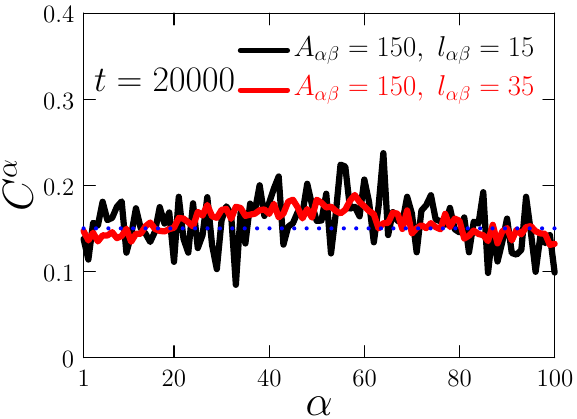}\label{hd2000}}\quad
\subfloat[]{\includegraphics[width=.48\textwidth]{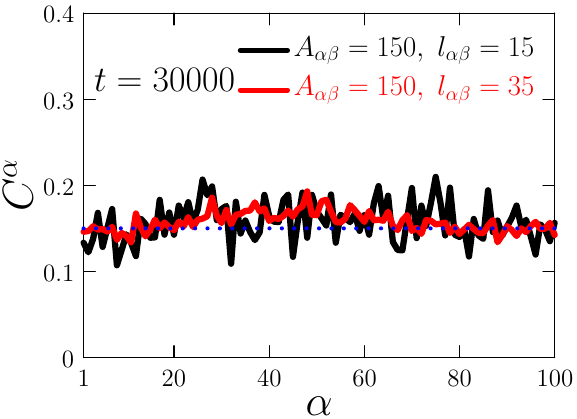}\label{hd3000}}\quad

\caption{Stochastic mass transportation process under RR2. The average concentration is 0.15 (outside of miscibility gap).}
\label{masstrans}
\end{figure}

\begin{figure}[ht!]
\centering
\includegraphics[width=.7\textwidth]{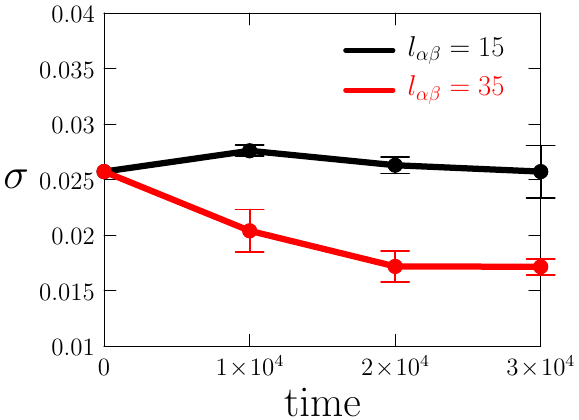}
\caption{Standard diviation of solute concentration at various time followed by Fig.\ref{masstrans}.}
\label{std_hd}
\end{figure}

Finally, we show the evolution of a concentration profile outside the miscibility gap, i.e., a simple mass transport process without driving force for phase separation. We use the RR2 approach but relaxing the constrain for a fixed volume. 
The initial concentration profile was sampled from a normal distribution with average concentration $c_0 = 0.15$ and standard deviation $\sigma = 25$. Figure~\ref{masstrans} presents concentration distribution profiles at four different times and different discretizations. The profiles for the different discretizations at every time step are closely overlapping each other. Furthermore, the standard deviation of elemental concentration
from the profiles shown in Fig.\ref{masstrans} is presented in Fig.\ref{std_hd}, in which differences in $\sigma$ ($\sim 0.01$)  are observed. Such differences stem from the different fluctuations. However, in both cases, the time evolution leads the system to a homogeneous state following the same kinetics. In this case, the system does not need to overcome any free energy barrier to reach equilibrium and hence the details of the fluctuations are less important.




\section{Discussion}\label{discussion}
The main goal of this work is to propose a new stochastic integration algorithm to solve the time evolution of the mesoscale phase field equations and to show that the fluctuations, naturally appearing from the algorithm, allow the system to overcome free energy barriers to reach equilibrium. To demonstrate such capability we have used a set of standard parameters taken from the literature, without focusing on replicating any specific result. Hence, none of the parameters or the expression for $f_{\rm chem}$ (Table~\ref{tab:msp} and Eq.~\eqref{f_density}) in our work is representative of a real material process. Yet, the following comments need to be stressed.  

\subsection{Stochastic noise}
The stochastic noise in SRSPF is coming from two sources. One is mentioned in both sections~\ref{intro} and~\ref{method}, that the residence time algorithm (RTA) integration scheme employed to integrate the phase field equations leads to statistical variations in the evolution, in contrast to deterministic approaches. The introduction of $\omega$ modifies the rates and therefore the timesteps. If the free energy wells are not harmonic, as it is the case here, a large absolute value of $\omega$ will result in deviations from equilibrium, as shown in Fig.~\ref{omega}.
The value of $\omega$ needs to be obtained from the details of some specific physical process. However, the larger the magnitude of $\omega$ the lower the timestep, and hence a value that balance both accuracy and computational efficiency would be optimal. 
\begin{figure}[ht!]
\centering
{\includegraphics[width=.48\textwidth]{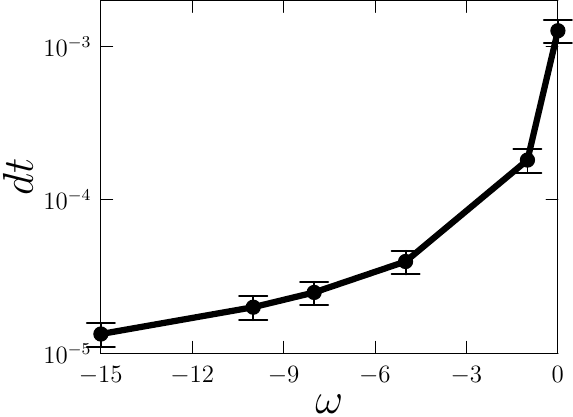}\label{dt_omega}}
\caption{Average time interval as a function of $\omega$.}
\label{dtomega}
\end{figure}
Figure ~\ref{dtomega} shows the relation between $\omega$ and the timestep. As $\omega$ increases so does the timestep in a larger than linear relation, which in terms of efficiency implies that the number of steps to reach a certain simulated time depends strongly on $\omega$. Controlling $\omega$ will have profound consequences on computational efficiency.



\subsection{Consistence with classical nucleation theory}
We have seen above that the results for the nucleation rate follow classical nucleation theory. Figure~\ref{nr_s0} shows a similar trend compared with Ref.\cite{PhysRevB.62.203} where a monotonic decrease of $\theta\left(\kappa\right)$ with increasing interface free energy (or $\kappa$) is highlighted. In Ref. \cite{PhysRevB.62.203}, $\theta\left(\sigma\right)$ has a cubic relationship with the interface free energy $\sigma$. However in our case, $\theta$ is a close to linear function of $\kappa$. As shown by Cahn \cite{Cahn1958}, $\kappa$ is a function of  $\sigma$ but also of $\Delta f_{\rm chem}(c)$. The differences in the definition of $\sigma$ and $\kappa$ are probably at the origin of this discrepancy. 

\subsection{Rescaling Rules}
Grid size effects in phase field methods have been described previously. Thermodynamic and kinetic parameters depend on the volume element. These dependencies have been shown to appear directly from the master equation. Brochant et al. \cite{RN1} derived the phase field equations coarse graining the master equation including size dependent noise. They computed the dependence of free energy and interface energy on size from independent Monte Carlo simulations. Their final phase field equations contain a Langevin term that provides the stochastic noise for the system to overcome free energy barriers. Furthermore, they showed that accounting for the volume element size in the model parameters leads to solutions independent of the discretization.
Garnier et al. \cite{RN5} showed that both homogeneous free energy and interface free energy depend on the discretization and develop expressions to account for such dependence. Furthermore, Garnier and Nastar \cite{RN5} developed what they call a coarse-grain kinetic Monte Carlo algorithm, similar in spirit to the algorithm proposed in this work but starting from the master equation with jump rates provided by atomic description while in the proposed model the rates are computed directly from the phase field equations.

The main idea for us to have section~\ref{rescale} is not to claim that the two rescaling rules may be applied, but to show that provided that the size dependence of the physical paramters are taken into account, the SRSPF should give the right kinetic evolution, as empirically shown above. In those previous works, \cite{RN1, RN5} they parameterized their models with lower scale methods providing the dependence to renormalize their equations in a similar fashion as proposed here.



\subsection{Computational efficiency}
Integrating the phase field equations stochastically following the residence time algorithm adds extra computational burden. The timesteps that we have observed in our simulations range from $10^{-5}$ to $10^{-3}$ time units, with each step taking $2\times10^{-4}$ wall clock seconds on average. To reach $1$ time unit we would need about $2$ wall clock seconds. One strategy that can be followed to improve the efficiency of the algorithm is to implement parallelization schemes \cite{MARTINEZ20111359, MARTINEZ20083804, Nandipati_2009}. Using a checkerboard method will avoid both causality and boundary errors and could potentially increase efficiency even in serial as the timestep increases and the need for updating the rates might decrease.

Once the number of nuclei reaches its maximum, a deterministic integration scheme could be used to accelerate the analysis of the growth regime that would not depend on thermal fluctuations 


\section{Conclusion}\label{conclusion}
We finalize the paper with the following most important conclusions:
\begin{enumerate}
\item We developed a stochastic solver for phase field models based on the residence-time algorithm, recasting the phase field equations into rate equations. This model naturally introduces fluctuations by stochastically sampling events, allowing for uphill particle fluxes. We have shown that these fluctuations indeed help the system overcome free energy barriers in the nucleation and growth domain of the phase diagram. 
\item The proposed algorithm complies with classical nucleation theory and gives an exponential nucleation rate depending on supersaturation and interface properties.
\item The fluctuations are discretization dependent and hence the precipitation kinetics depend on the grid size. However, we have empirically shown that if the size dependent phase field parameters are provided, the algorithm will accurately predict the kinetic evolution. 
\item Parallelization schemes and combinations with deterministic algorithms could be developed to increase the computational efficiency.
\end{enumerate}

\section*{CREdiT authorship contribution statement}
Q. Yu: Conceptualization, Methodology, Software, Validation, Formal analysis, Investigation, Writing - Original Draft, Writing - Review \& Editing, Visualization. N. Julian: Conceptualization, Methodology, Software, Validation, Formal analysis, Investigation, Writing - Original Draft, Writing - Review \& Editing, Visualization. J. Marian: Conceptualization, Methodology, Software, Validation, Formal analysis, Investigation, Writing - Original Draft, Writing - Review \& Editing, Visualization. E. Martinez: Conceptualization, Methodology, Writing - Review \& Editing, Project administration, Funding acquisition.

\section*{Acknowledgements}
Discussions with M. Nastar, F. Soisson and A. Finel are gratefully acknowledge. Q.Y. and E.M. were supported by the U.S.
Department of Energy, Office of Science, FES program
under Award Number DE-SC-00272037.

Additionally, this material is based on work supported by the National Science Foundation under Grant Nos. MRI\# 2024205, MRI\# 1725573, and CRI\# 2010270 for allotment of compute time on the Clemson University Palmetto Cluster.


\begin{thebibliography}{10}
\expandafter\ifx\csname url\endcsname\relax
  \def\url#1{\texttt{#1}}\fi
\expandafter\ifx\csname urlprefix\endcsname\relax\def\urlprefix{URL }\fi
\expandafter\ifx\csname href\endcsname\relax
  \def\href#1#2{#2} \def\path#1{#1}\fi

\bibitem{Cahn1958}
J.~W. Cahn, J.~E. Hilliard, Free energy of a nonuniform system. i. interfacial
  free energy, The Journal of Chemical Physics 28~(2) (1958) 258--267.

\bibitem{refId0}
{CAHN, J. W.}, {ALLEN, S. M.}, A microscopic theory for domain wall motion and
  its experimental verification in fe-al alloy domain growth kinetics, Le
  Journal de Physique Colloques 38~(C7) (1977) C7--51--C7--54.

\bibitem{RN46}
M.~R. Tonks, L.~K. Aagesen, The phase field method: Mesoscale simulation aiding
  material discovery, Annual Review of Materials Research 49~(1) (2019)
  79--102.

\bibitem{RN80}
C.-H. Chen, E.~Bouchbinder, A.~Karma, Instability in dynamic fracture and the
  failure of the classical theory of cracks, Nature Physics 13~(12) (2017)
  1186--1190.

\bibitem{YANG20191}
Y.~Yang, J.~Wang, S.~Zhou, T.~Huang, Design of a novel coaxial eccentric
  indexing cam mechanism, Mechanism and Machine Theory 132 (2019) 1--12.

\bibitem{MCCUE201610}
I.~McCue, B.~Gaskey, P.-A. Geslin, A.~Karma, J.~Erlebacher, Kinetics and
  morphological evolution of liquid metal dealloying, Acta Materialia 115
  (2016) 10--23.

\bibitem{RN63}
M.~Plapp, Remarks on some open problems in phase-field modelling of
  solidification, Philosophical Magazine 91~(1) (2011) 25--44.

\bibitem{WANG19982983}
Y.~Wang, D.~Banerjee, C.~Su, A.~Khachaturyan, Field kinetic model and computer
  simulation of precipitation of l12 ordered intermetallics from f.c.c. solid
  solution, Acta Materialia 46~(9) (1998) 2983--3001.

\bibitem{Karma_1999}
A.~Karma, W.-J. Rappel, Phase-field model of dendritic sidebranching with
  thermal noise, Physical Review E 60~(4) (1999) 3614--3625.

\bibitem{RN54}
L.~Granasy, T.~Borzsonyi, T.~Pusztai, Crystal nucleation and growth in binary
  phase-field theory, Journal of Crystal Growth 237-239 (2002) 1813--1817.

\bibitem{RN142797}
Q.~Yu, M.~Reyes, N.~Shah, J.~Marian, Kinetic model of incipient hydride
  formation in zr clad under dynamic oxide growth conditions, Materials 13~(5)
  (2020) 1088.

\bibitem{RN142998}
Q.~Yu, M.~J. Simmonds, R.~Doerner, G.~R. Tynan, L.~Yang, B.~D. Wirth,
  J.~Marian, Understanding hydrogen retention in damaged tungsten using
  experimentally-guided models of complex multispecies evolution, Nuclear
  Fusion 60~(9) (2020) 096003.

\bibitem{RN47}
D.~Tourret, H.~Liu, J.~Llorca, Phase-field modeling of microstructure
  evolution: Recent applications, perspectives and challenges, Progress in
  Materials Science 123 (2022) 100810.

\bibitem{RN62}
L.~Granasy, G.~I. Tath, J.~A. Warren, F.~Podmaniczky, G.~Tegze, L.~Ratkai,
  T.~Pusztai, Phase-field modeling of crystal nucleation in undercooled liquids
  ? a review, Progress in Materials Science 106 (2019) 100569.

\bibitem{RN64}
A.~M. Jokisaari, P.~W. Voorhees, J.~E. Guyer, J.~Warren, O.~G. Heinonen,
  Benchmark problems for numerical implementations of phase field models,
  Computational Materials Science 126 (2017) 139--151.

\bibitem{RN142622}
D.~T. Gillespie, A general method for numerically simulating the stochastic
  time evolution of coupled chemical reactions, Journal of Computational
  Physics 22~(4) (1976) 403--434.

\bibitem{RN22}
J.~Marian, V.~V. Bulatov, Stochastic cluster dynamics method for simulations of
  multispecies irradiation damage accumulation, Journal of Nuclear Materials
  415~(1) (2011) 84--95.

\bibitem{RN20}
J.~Marian, T.~L. Hoang, Modeling fast neutron irradiation damage accumulation
  in tungsten, Journal of Nuclear Materials 429~(1?3) (2012) 293--297.

\bibitem{PhysRevB.62.203}
F.~Soisson, G.~Martin, Monte carlo simulations of the decomposition of
  metastable solid solutions: Transient and steady-state nucleation kinetics,
  Phys. Rev. B 62 (2000) 203--214.

\bibitem{RN1}
Q.~Bronchart, Y.~Le~Bouar, A.~Finel, New coarse-grained derivation of a phase
  field model for precipitation, Physical Review Letters 100~(1) (2008) 015702,
  pRL.

\bibitem{RN5}
T.~Garnier, M.~Nastar, Coarse-grained kinetic monte carlo simulation of
  diffusion in alloys, Physical Review B 88~(13) (2013) 134207, pRB.

\bibitem{MARTINEZ20111359}
E.~Martínez, P.~Monasterio, J.~Marian, Billion-atom synchronous parallel
  kinetic monte carlo simulations of critical 3d ising systems, Journal of
  Computational Physics 230~(4) (2011) 1359--1369.

\bibitem{MARTINEZ20083804}
E.~Martínez, J.~Marian, M.~Kalos, J.~Perlado, Synchronous parallel kinetic
  monte carlo for continuum diffusion-reaction systems, Journal of
  Computational Physics 227~(8) (2008) 3804--3823.

\bibitem{Nandipati_2009}
G.~Nandipati, Y.~Shim, J.~G. Amar, A.~Karim, A.~Kara, T.~S. Rahman, O.~Trushin,
  Parallel kinetic monte carlo simulations of ag(111) island coarsening using a
  large database, Journal of Physics: Condensed Matter 21~(8) (2009) 084214.

\end{thebibliography}

\end{document}